\documentclass[onecollarge]{svjour2}

\usepackage{amsbsy,amssymb,amsmath,bm}
\usepackage{graphicx,subfigure}
\usepackage[authoryear,round]{natbib}

\begin{document}

\title{Decay of isotropic flow and anisotropic flow with rotation or magnetic field or both in a weakly nonlinear regime}
\author{Xing Wei}
\institute{\at Institute of Natural Sciences and Department of Physics and Astronomy, Shanghai Jiao Tong University, Shanghai 200240, China, \email{xing.wei@sjtu.edu.cn} \\
           \at Princeton University Observatory, Princeton NJ 08544, USA, \email{xingwei@astro.princeton.edu}}
\date{Received: date / Accepted: date}

\maketitle

\begin{abstract}
We investigate numerically the decay of isotropic, rotating, magnetohydrodynamic (MHD), and rotating MHD flows in a periodic box. The Reynolds number $Re$ defined with the box size and the initial velocity is $100$ at which the flows are in a weakly nonlinear regime, i.e. not laminar but far away from the fully turbulent state. The decay of isotropic flow has two stages, the first stage for the development of small scales and the second stage for the viscous dissipation. In the rapidly rotating flow, fast rotation induces the inertial wave and causes the large-scale structure to inhibit the development of the first stage and retard the flow decay. In the MHD flow, the imposed field also causes the large-scale structure but facilitates the flow decay in the first stage because of the energy conversion from flow to magnetic field. Magnetic Reynolds number $Rm$ is important for the dynamics of the MHD flow, namely a high $Rm$ induces the Alfv\'en wave but a low $Rm$ cannot. In the rotating MHD flow, slower rotation tends to convert more kinetic energy to magnetic energy. The orientation between the rotational and magnetic axes is important for the dynamics of the rotating MHD flow, namely the energy conversion is more efficient and the stronger wave is induced when the two axes are not parallel than when they are parallel.
\end{abstract}

\section{Introduction}\label{sec:intro}

Rotation and magnetic field play important roles in the engineering, geophysical and astrophysical fluid motions. Rotation causes the Coriolis force in the rotating frame and magnetic field causes the Lorentz force. These body forces act as restoring forces to induce the internal waves that transport energy and angular momentum in the fluid interior. Rotation induces the inertial wave \citep{greenspan}, magnetic field induces the Alfv\'en wave \citep{davidson1}, and the combined effect of rotation and magnetic field induces magneto-Coriolis wave \citep{moffatt}. The propagation of these waves leads to the anisotropy of flows and the large-scale structure forms along the rotational axis in the rapidly rotating flow or along the magnetic field lines in the MHD flow. The decay of the rotating turbulence and the MHD turbulence has been studied, for example, the experimental work of \citet{sreenivasan,staplehurst}, etc., the analytical work of \citet{moffatt1,goldreich,davidson3}, etc., and the numerical work of \citet{thiele,okamoto,lee,teitelbaum,wan}, etc. It has been already pointed out that the decay rate depends on the initial flow in the isotropic turbulence \citep{ishida} as well as in the anisotropic turbulence \citep{davidson3}. Moreover, in the MHD turbulence the decay rate also depends on the ratio of various time scales, namely the lack of universality as suggested by \citet{moffatt1,lee}, etc. We will discuss this in \S\ref{sec:mhd}. However, the rotating MHD turbulence is not extensively studied, especially how the different orientations of the rotational and magnetic axes will influence the decay of the rotating MHD turbulence. On the other hand, the weakly nonlinear regime is not well studied. Through our moderate-scale numerical calculations in the weakly nonlinear regime, we will answer the question whether the isotropic and anisotropic flows in the weakly nonlinear regime exhibit the similar behaviour to the turbulent flows or they are quite different from the turbulence. In addition, we will study the influence of the orientation between rotation and magnetic field on the rotating MHD flow in the weakly nonlinear regime, which is of help to understanding the rotating MHD turbulence.

In this paper we will numerically study the decay of unforced flows in the presence of rotation or magnetic field or both. The geometry is a periodic box, i.e. unbounded flows. The Reynolds number $Re$ is defined with the box size and the initial velocity. It should be noted that our Reynolds number is different from the conventional Reynolds number $R_\lambda$ defined with the turbulent fluctuating velocity and the Taylor micro-scale $\lambda$ which is often used in the study of turbulence. We choose $Re=100$ in almost all parts of this paper, except that in \S\ref{sec:iso} we will calculate $Re=200$ for comparison with $Re=100$. At such the low $Re=100$ (note: not $R_\lambda$ but $Re$ defined with the large scale of box size and initial velocity), the flows are in the weakly nonlinear regime, i.e. they are not laminar but far away from the fully turbulent state. In the next sections we will study step-by-step the isotropic flow, the rotating flow, the MHD flow and the rotating MHD flow.

\section{Isotropic flow}\label{sec:iso}
Before studying the anisotropic flow in the presence of rotation or magnetic field, we begin with isotropic flow. We study in the Cartesian coordinate system $(x_1,x_2,x_3)$. The computational geometry is a cube with its size to be $2\pi l$ and subject to the periodic boundary condition in the $x_1$, $x_2$ and $x_3$ directions. This periodic box can be considered as a small piece of region taken out of the flow interior. We give an initial velocity $\bm u_0$ and then numerically calculate the governing equation of fluid motion for the decay problem. The dimensionless Navier-Stokes equation governing incompressible fluid motion reads
\begin{equation}
\frac{\partial\bm u}{\partial t}+\bm u\cdot\bm\nabla\bm u=-\bm\nabla p+\frac{1}{Re}\nabla^2\bm u,
\end{equation}
where length is normalised with $l$, velocity with the volume-averaged initial velocity $\bar u_0=\sqrt{\int u_0^2dV/V}$, time with $l/\bar u_0$ and pressure with $\rho\bar u_0^2$ where $\rho$ is fluid density. Reynolds number is defined as
\begin{equation}
Re=\frac{\bar u_0 l}{\nu}
\end{equation}
where $\nu$ is viscosity. 

To satisfy the incompressible condition $\bm\nabla\cdot\bm u_0$, the initial velocity is simply given to be
\begin{equation}
u_{01}=\sin(k_0x_2)\sin(k_0x_3),u_{02}=\sin(k_0x_3)\sin(k_0x_1),u_{03}=\sin(k_0x_1)\sin(k_0x_2),
\end{equation}
where $k_0$ is the initial wave number and measures the length scale of initial flow. It should be noted that the dynamics of decaying turbulence depends on the initial condition as pointed out in \citet{ishida}, namely the linear momentum conservation or the angular momentum conservation of the initial flow. However, we are not concerned with this subtle point in this short paper. On the other hand, our initial flow for the study in the weakly nonlinear regime is on the large scale but in the study of fully nonlinear turbulence the initial flow is on the small scale, say, $k_0$ up to 80 in \citet{ishida}.

We output the volume-averaged kinetic energy and viscous dissipation. With the periodic boundary condition, viscous dissipation is equal to enstrophy multiplied by viscosity. We normalise kinetic energy with $u_0^2$ and viscous dissipation with $u_0^3/l$, and so their dimensionless expressions are respectively
\begin{equation}
\frac{1}{V}\int\frac{1}{2}|\bm u|^2dV \hspace{3mm}{\rm and}\hspace{3mm} \frac{1}{Re}\frac{1}{V}\int|\bm\nabla\times\bm u|^2dV.
\end{equation}

The numerical method is the standard pseudo-spectral algorithm. Taking the divergence of Navier-Stokes equation leads to Poisson's equation of pressure
\begin{equation}
\nabla^2 p=\bm\nabla\cdot\bm f,
\end{equation}
where $\bm f=-\bm u\cdot\bm\nabla\bm u$ is the inertial force. In the next sections $\bm f$ involves the other body forces, i.e. the Coriolis force in the rotating flow and the Lorentz force in the MHD flow. Because of the periodic boundary condition we assume that velocity and pressure are expressed as
\begin{equation}
\bm u=\hat{\bm u}(t)e^{i\bm k\cdot\bm x} \hspace{3mm}{\rm and}\hspace{3mm} p=\hat{p}(t)e^{i\bm k\cdot\bm x},
\end{equation}
where hat denotes spectral space and $\bm k$ is wave vector. Solving the pressure equation in spectral space and substituting into Navier-Stokes equation, we derive
\begin{equation}
\left(\frac{d}{dt}+\frac{k^2}{Re}\right)\hat{\bm u}=-\frac{\bm k\cdot\hat{\bm f}}{k^2}\bm k+\hat{\bm f},
\end{equation}
where the viscous term is treated implicitly. The nonlinear term $\bm f$ is calculated in physical space and $\hat{\bm f}$ in spectral space is calculated by fast Fourier transform. Time is stepped forward with the prediction-correction method. The second-order Runge-Kutta method is used for time stepping. The resolution as high as $128^3$ and the time step as small as $10^{-2}$ are used. The fast Fourier transform is used and the dealiase is satisfied in the discrete Fourier transform.

\begin{figure}
\centering
\includegraphics[scale=0.7]{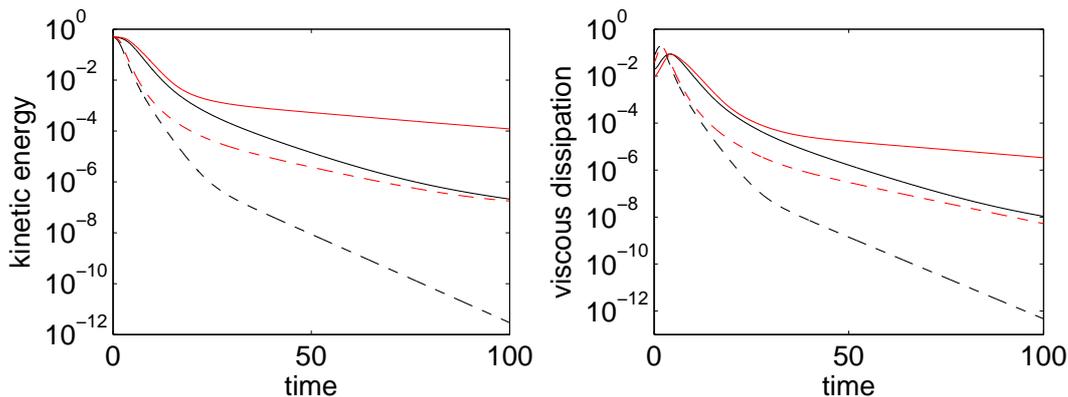}
\caption{Isotropic flow. The time evolution of volume-averaged kinetic energy and viscous dissipation. Black lines denote $Re=100$ and red lines $Re=200$. Solid lines denote $k_0=1$ and dashed lines $k_0=2$.}\label{fig:iso1}
\end{figure}

We calculate the four isotropic flows with the combination of $Re=(100, 200)$ and $k_0=(1,2)$. Figure \ref{fig:iso1} shows the time evolution of the volume-averaged kinetic energy and viscous dissipation. Time is integrated until the order of viscous time scale $t_\nu=Re=O(10^2)$. The left panel indicates that kinetic energy decays for two stages. In the first stage flow develops to small scales through the nonlinear coupling, and then in the second stage kinetic energy decays through viscosity. The right panel verifies these two stages. In the first stage enstrophy increases to its peak because of formation of small scales, and then in the second stage enstrophy decays through viscosity. The flows with higher $k_0$ have an earlier first stage because small scales develop earlier with higher $k_0$. These two stages in the weakly nonlinear flow are similar to the cascade and dissipation stages in the fully turbulent flow \citep{davidson2,ishida}. It is not surprising that in both the first and the second stages higher $Re$ corresponds to slower decay because of smaller viscosity and higher $k_0$ corresponds to faster decay because of smaller scales. We will see in the next sections that the energy cascade in the anisotropic flows with fast rotation or a strong magnetic field will not have the two distinct stages as in the isotropic flow because of the two dimensionalization.

\begin{figure}
\centering
\includegraphics[scale=0.35]{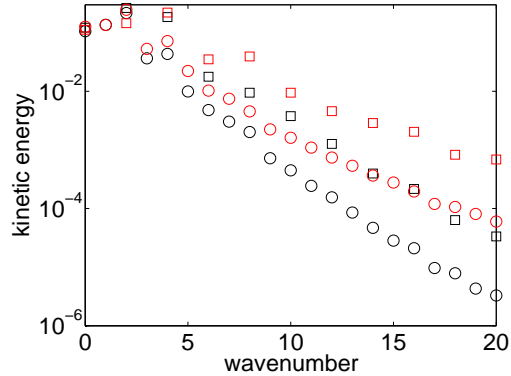}
\caption{Isotropic flow. The spectrum of kinetic energy at the snapshot when enstrophy reaches its peak value. Black symbols denote $Re=100$ and red symbols $Re=200$. Circle symbols denote $k_0=1$ and square symbols $k_0=2$. Spectrum is truncated at $k=20$ for better view.}\label{fig:iso2}
\end{figure}

Figure \ref{fig:iso2} shows the spectrum of kinetic energy at the snapshot when enstrophy reaches its peak value (for isotropic flow the spectra in all the three directions are identical). It indicates that higher $Re$ or $k_0$ corresponds to wider spectrum, because higher $Re$ leads to stronger nonlinearity and higher $k_0$ leads to coupling with wider gap. The flow at $Re=100$ and with $k_0=1$ shows a visible energy spectrum, which suggests that the flow is in the weakly nonlinear regime.

After this simple investigation on $Re$ and $k_0$ for isotropic flow, we will use these two parameters, $Re=100$ and $k_0=1$, to calculate the weakly nonlinear flows in the next sections for the studies of anisotropic flows.

\section{Rotating flow}\label{sec:rot}

In this section we study the decay of the rotating flow. Suppose that the cube rotates about the $x_3$ axis at a constant angular velocity $\bm\Omega=\Omega \hat{\bm x}_3$, hat denoting unit vector. In the frame rotating at $\bm\Omega$ the dimensionless Navier-Stokes equation reads
\begin{equation}
\frac{\partial\bm u}{\partial t}+\bm u\cdot\bm\nabla\bm u=-\bm\nabla p+\frac{1}{Re}\nabla^2\bm u+\frac{1}{Ro}2\bm u\times\hat{\bm x}_3,
\end{equation}
where the curl-free centrifugal force is absorbed into pressure gradient. The Rossby number
\begin{equation}
Ro=\frac{\bar u_0}{\Omega l}
\end{equation}
measures the ratio of the inertial force to the Coriolis force. In the rapidly rotating flow $Ro$ is lower than unity and the Coriolis force wins, whereas in the slowly rotating flow $Ro$ is higher than unity and inertial force wins.

\begin{figure}
\centering
\includegraphics[scale=0.7]{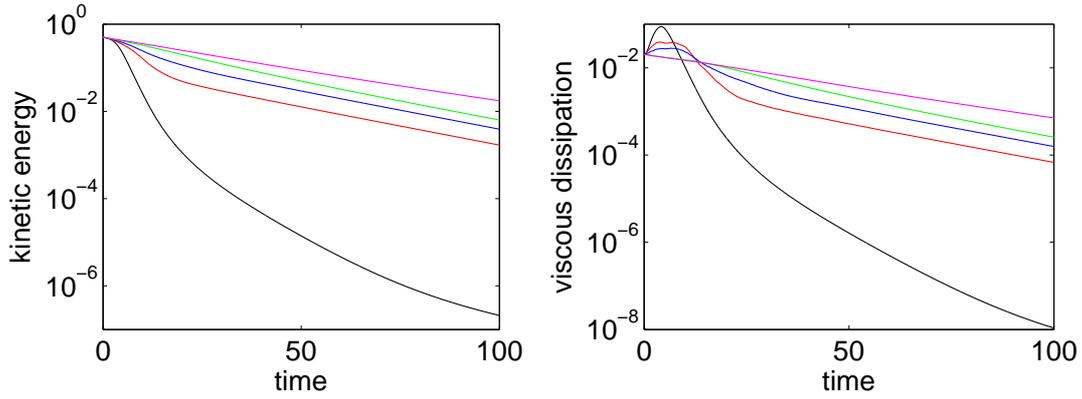}
\caption{Rotating flow. The time evolution of volume-averaged kinetic energy and viscous dissipation. Black lines denote isotropic flow. Red, blue, green and magenta lines denote respectively $Ro=1.0$, $0.5$, $0.2$ and $0.1$.}\label{fig:rot1}
\end{figure}

We calculate the rotating flows with $Ro$ decreasing from $1$ to $0.1$, for which the Coriolis force wins and rotation becomes faster and faster. Figure \ref{fig:rot1} shows the time evolution of volume-averaged kinetic energy and viscous dissipation. The left panel indicates that the rotating flows decay much more slowly than the isotropic flow, and faster rotation at lower $Ro$ corresponds to slower decay in the first stage, but the decay rates of the rotating flows at different $Ro$'s do not differ too much in the second stage. This suggests that rotation takes its effect on flow decay mainly in the first stage while viscosity takes its effect in the second stage. The right panel indicates that the large-scale structure forms at low $Ro$. Enstrophy at $Ro=1.0$ and $0.5$ initially increases to the peak and then decays, which is similar to isotropic flow. $Ro=0.5$ corresponds to a lower peak than $Ro=1.0$ because flow at lower $Ro$ has larger scale. At $Ro=0.2$ and $0.1$ enstrophy does not have the initial increase, because flows at sufficiently low $Ro$ do not develop to small scales. Therefore, fast rotation inhibits the first stage for development of small scales.

\begin{figure}
\centering
\includegraphics[scale=0.35]{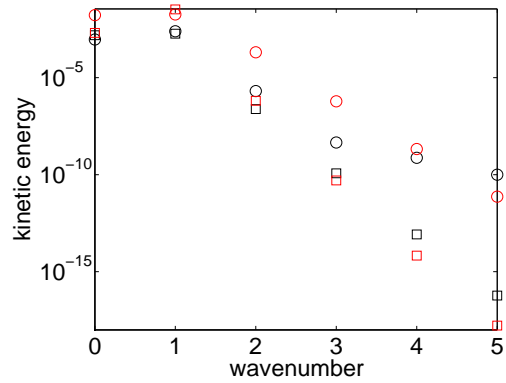}
\caption{Rotating flow. The spectrum of kinetic energy at time$=100$. Black symbols denote $Ro=1.0$ and red symbols $Ro=0.1$. Circle symbols denote $k_1$ the spectrum and square symbols the $k_3$ spectrum. Spectrum is truncated at $k=5$ for better view.}\label{fig:rot2}
\end{figure}

\begin{figure}
\centering
\includegraphics[scale=0.7]{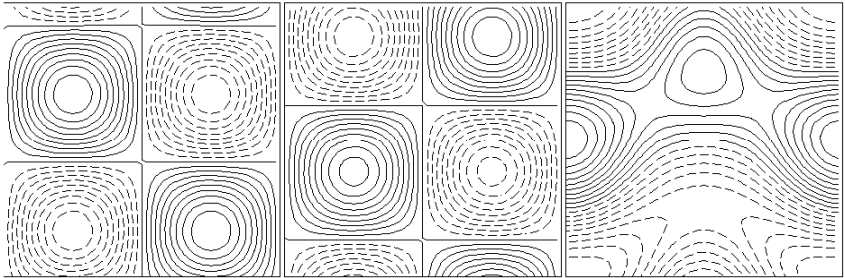}
\caption{Rotating flow. The contours of the three components of velocity in the $x_1-x_3$ plane. From left to right the three panels are $u_1$, $u_2$ and $u_3$. $Ro=0.1$.}\label{fig:cont}
\end{figure}

Figure \ref{fig:rot2} shows the $k_1$ and $k_3$ spectra of kinetic energy at time$=100$ for $Ro=1.0$ (the slowest rotation) and $0.1$ (the fastest rotation). The $k_2$ spectrum is the same as the $k_1$ spectrum. At the slow rotation the $k_1$ and $k_3$ spectra do not differ too much and both concentrate on the first two modes, namely the flow is almost isotropic. With the fast rotation the $k_3$ spectrum in the rotational direction concentrates on the $k_3=1$ mode but the $k_1$ spectrum on the first two modes, namely the flow is anisotropic. Moreover, the difference between the two energies contained in the $k_1$ and the $k_3$ modes for the same wavenumber (the difference between circle and square symbols for the same wavenumber) is larger with the faster rotation than the slower rotation. Therefore, the faster rotation leads to the more anisotropy of flow. Figure \ref{fig:cont} shows the contours of the three components of velocity in the $x_1-x_3$ plane. $u_1$ and $u_2$ have similar structure whereas $u_3$ has quite different structure. It can interpreted by the equation of motion. The Coriolis force $(1/Ro)2\bm u\times\hat{\bm x}_3$ has no component in the $x_3$ direction such that the equation of $u_3$ has no Coriolis term. Therefore, the energy contained in the three components are very different.

\begin{figure}
\centering
\includegraphics[scale=0.7]{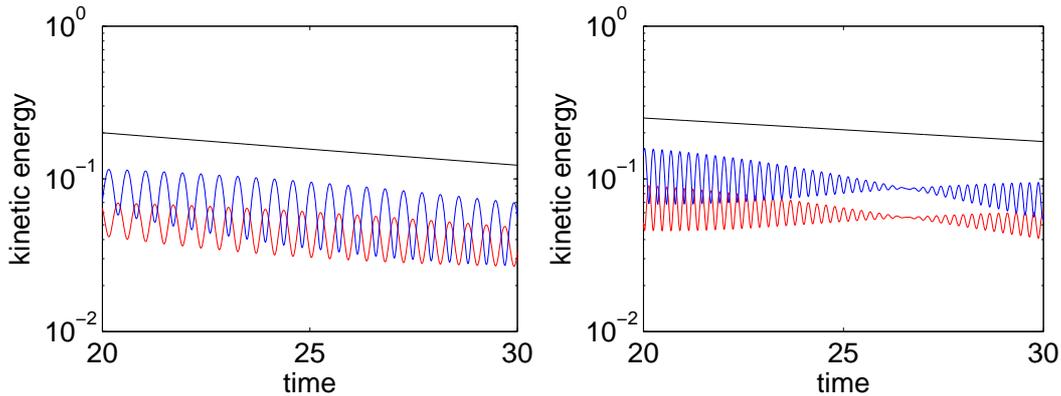}
\caption{Rotating flow. The time evolution of volume-averaged kinetic energy to show inertial wave. Black line denotes the total kinetic energy, red line the kinetic energy in the $x_1$ direction and blue line the kinetic energy in the $x_3$ direction. Left panel is for $Ro=0.2$ and the oscillation period is $0.45$. Right panel is for $Ro=0.1$ and the oscillation period is $0.22$. Time is truncated at $t=30$ for better view.}\label{fig:rot3}
\end{figure}

Interestingly, in the rotating flow there exist oscillations of kinetic energy in the three directions. Figure \ref{fig:rot3} shows the time evolution of volume-averaged total kinetic energy, and kinetic energy in the $x_1$ direction (i.e. $1/V\int u_1^2/2dV$) and in the $x_3$ direction (i.e. $1/V\int u_3^2/2dV$). Kinetic energy in the $x_2$ direction is the same as in the $x_1$ direction. The oscillation period is $0.45$ for $Ro=0.2$ shown in the left panel and $0.22$ for $Ro=0.1$ shown in the right panel. The frequency of inertial wave is $\omega=2\Omega\cos\theta$ where $\theta$ is the angle between the wave vector and the rotational axis. With our normalisation, the period of inertial wave is $\pi Ro/\cos\theta$, namely it is proportional to $Ro$. The fact that the two periods in the figure are proportional to $Ro$ suggests that these oscillations arise from inertial wave. Furthermore, we can calculate the angle $\theta$. The period of kinetic energy should be half of the period of inertial wave, e.g. $u_1^2\propto e^{2i\omega t}$, and the substitution of numerical values gives $\theta=45^\circ$. In addition, comparison between the left and right panels indicates that faster rotation causes more difference of kinetic energy between the two directions. Again, faster rotation leads to more anisotropy.

\section{MHD flow}\label{sec:mhd}

In this section we study decay of MHD flow. It is well known that the propagation of Alfv\'en wave along magnetic field lines will lead to anisotropy, e.g. \citet{reddy} numerically studied the strong anisotropy in the forced MHD turbulence. We assume that a uniform magnetic field is imposed in the $x_3$ direction. The assumption that the imposed field is uniform is valid for the situation that the length scale of variation of imposed field is much larger than that of flow and induced field, i.e. the local WKB approximation. We decompose the total magnetic field into the imposed field $\bm B_0$ and the induced field $\bm b$. The dimensionless Navier-Stokes equation then reads
\begin{equation}
\frac{\partial\bm u}{\partial t}+\bm u\cdot\bm\nabla\bm u=-\bm\nabla p+\frac{1}{Re}\nabla^2\bm u+V_A^2(\bm\nabla\times\bm b)\times(\hat{\bm x}_3+\bm b),
\end{equation}
where magnetic field is normalised with $B_0$. The dimensionless Alfv\'en speed
\begin{equation}
V_A=\frac{B_0}{\sqrt{\rho\mu}\bar u_0}
\end{equation}
measures the strength of imposed field relative to the initial flow. The magnetic induction equation reads
\begin{equation}
\frac{\partial\bm b}{\partial t}=\bm\nabla\times\left(\bm u\times(\hat{\bm x}_3+\bm b)\right)+\frac{1}{Rm}\nabla^2\bm b.
\end{equation}
The magnetic Reynolds number
\begin{equation}
Rm=\frac{\bar u_0 l}{\eta}
\end{equation}
measures the strength of induction effect against magnetic diffusion. There are two additional parameters in MHD flow, $V_A$ and $Rm$. The imposed field is measured by $V_A$ and the induced field by $Rm$. 

In addition to the $V_A$ and $Rm$ there are some other dimensionless parameters in the MHD flow and they are the ratios between different time scales. The damping time of Alfv\'en wave $t_d=\eta/\widetilde{V_A}^2$ where $\widetilde{V_A}=B_0/\sqrt{\rho\mu}$ is the dimensional Alfv\'en speed is called the magnetic damping time, and it should be noted that $t_d$ is different from the magnetic diffusion time $t_\eta=l^2/\eta$. Firstly we define the interaction parameter by the ratio of $l/\bar u_0$ to $t_d$
\begin{equation}
N=\frac{l/\bar u_0}{t_d}=\frac{lB_0^2}{\rho\mu\eta\bar u_0}.
\end{equation}
The interaction parameter is often used in the analysis of low $Rm$ MHD flow \citep{davidson1}. Then we define the Lundquist number as the ratio of the magnetic diffusion time $t_\eta$ to the Alfv\'enic time $l/\widetilde{V_A}$
\begin{equation}
S=\frac{l^2/\eta}{l/V_A}=\frac{l\widetilde{V_A}}{\eta}=\frac{lB_0}{\sqrt{\rho\mu}\eta}.
\end{equation}
The Lundquist number is often used in the analysis of MHD turbulence \citep{moffatt1}. $Rm$, $N$ and $S$ are related through
\begin{equation}
S=(RmN)^{1/2}.
\end{equation}
It is well known that the decay of MHD turbulence has the lack of universality, namely the decay rate depends on the strength of imposed field which can be measured by $V_A$, $N$ or $S$. For example, in \citep{moffatt1} the decay rate depends on $\zeta=\bm B_0\cdot\bm k/(\sqrt{\rho\mu}\eta k^2)$ where $\bm k$ is the wave vector of Fourier components of velocity.

Besides the kinetic energy and viscous dissipation in the hydrodynamic flow, in the MHD flow we need to output the magnetic energy and Ohmic dissipation of induced field. The former is normalised with $\bar u_0^2$ and the latter with $\bar u_0^3/l$, and so their dimensionless expressions are respectively
\begin{equation}
V_A^2\frac{1}{V}\int\frac{1}{2}|\bm b|^2dV \hspace{3mm}{\rm and}\hspace{3mm} \frac{V_A^2}{Rm}\frac{1}{V}\int|\bm\nabla\times\bm b|^2dV.
\end{equation}

\begin{figure}
\centering
\includegraphics[scale=0.75]{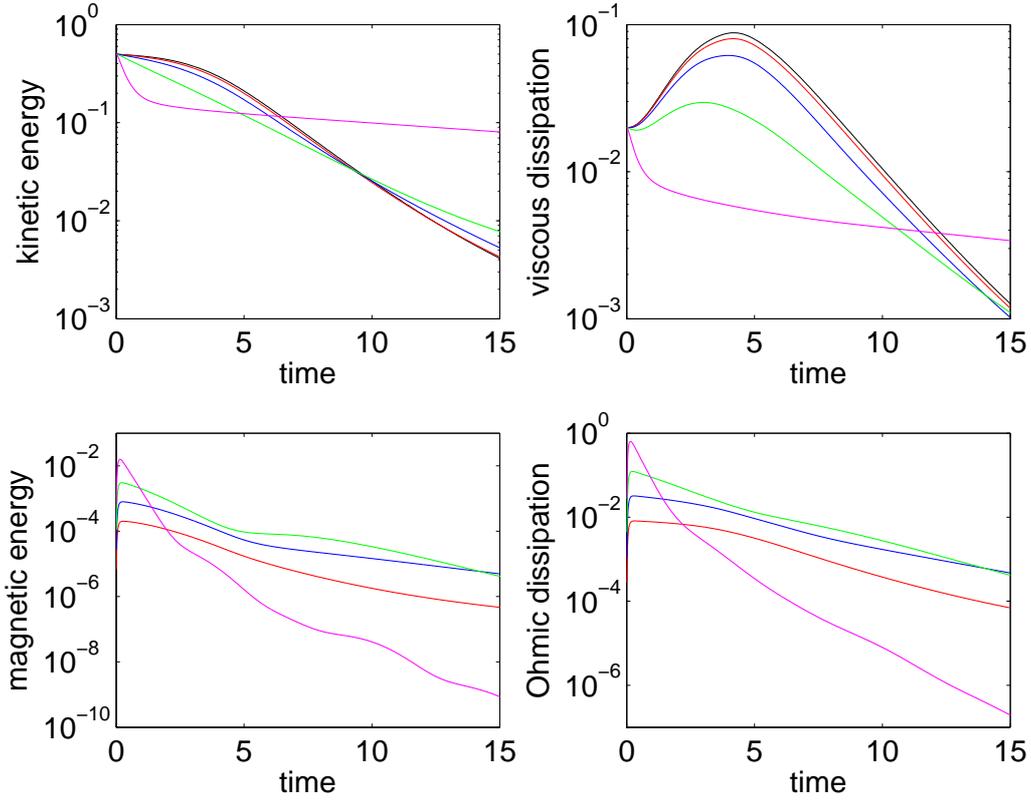}
\caption{MHD flow. The time evolution of volume-averaged kinetic energy, viscous dissipation, and magnetic energy and Ohmic dissipation of induced field. $Rm=0.1$. Black lines denote isotropic flow. Red, blue, green and magenta lines denote respectively $V_A=0.5$, $1$, $2$ and $5$. Time is truncated at $t=15$ for better view.}\label{fig:mhd1}
\end{figure}

Firstly we keep $Rm=0.1$ for a weak induced field and vary $V_A$ from $0.5$ to $5$ to study the effect of imposed field. Figure \ref{fig:mhd1} shows the time evolution of volume-averaged kinetic energy, viscous dissipation, and magnetic energy and Ohmic dissipation of induced field. The upper-left panel indicates that in the first stage the MHD flows decay faster than the isotropic flow and a stronger imposed field leads to faster decay. The upper-right panel indicates that in the first stage a stronger imposed field leads to a larger scale structure, and particularly, with the strongest imposed field the enstrophy does not increase but decrease. It seems opposite to the situation of the rotating flow, in which a larger scale structure with faster rotation leads to slower decay. To interpret this, we need to refer to the energy equation of MHD flow
\begin{align}
& -\frac{d}{dt}\mbox{(kinetic energy + magnetic energy of imposed field (equal to $V_A^2$)} \nonumber\\
& \hspace{32.5mm}\mbox{+ magnetic energy of induced field)} \nonumber\\
& =\mbox{viscous dissipation + Ohmic dissipation of induced field}.
\end{align}
This equation suggests that the kinetic and magnetic energies can be converted to each other due to the work done by Lorentz force. Then the two bottom panels give the reason for this discrepancy between rotating and MHD flows. This is because kinetic energy is converted to magnetic energy which is damped through Ohmic dissipation. Comparison between the upper-right and bottom-right panels indicates that Ohmic dissipation wins out viscous dissipation with the strength of imposed field increasing.

\begin{figure}
\centering
\includegraphics[scale=0.7]{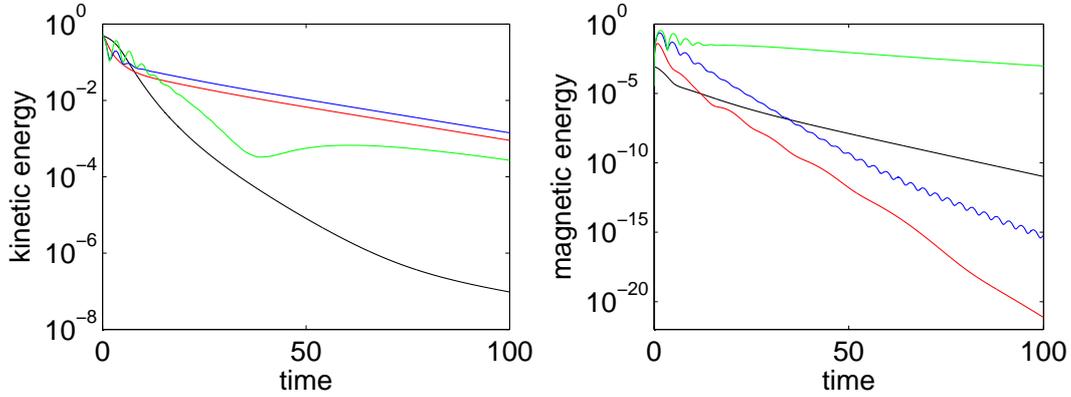}
\caption{MHD flow. The time evolution of volume-averaged kinetic energy and magnetic energy of induced field. $V_A=1$. Black, red, blue and green lines denote respectively $Rm=0.1$, $1$, $10$ and $100$.}\label{fig:mhd2}
\end{figure}

Next we keep $V_A=1$ and vary $Rm$ from $0.1$ to $100$ to study the effect of induced field. Figure \ref{fig:mhd2} shows the time evolution of volume-averaged kinetic energy and magnetic energy of induced field. The left panel indicates that kinetic energy at the lowest $Rm=0.1$ and the highest $Rm=100$ decays faster than at the other two intermediate $Rm$'s, and this is not straightforward to understand. The right panel indicates that the highest $Rm$ corresponds to the strongest induced field, and this is not surprising.

\begin{figure}
\centering
\includegraphics[scale=0.7]{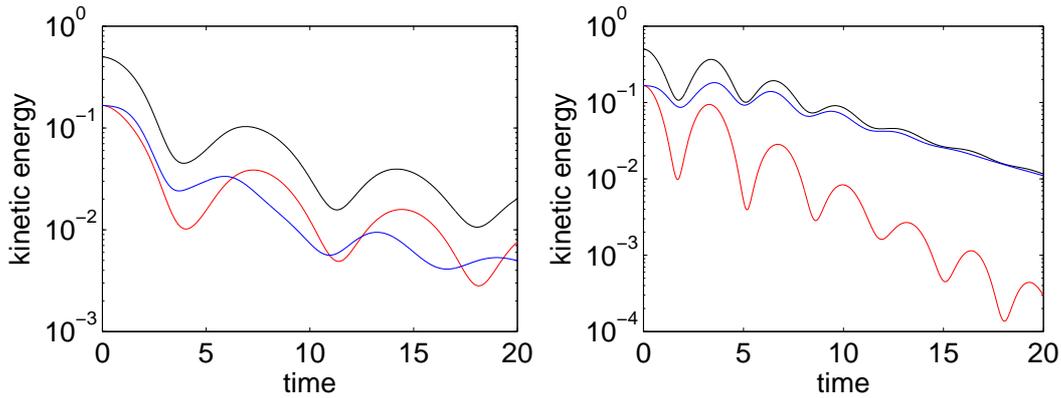}
\caption{MHD flow. The time evolution of volume-averaged kinetic energy to show Alfv\'en wave. $Rm=100$. Black line denotes the total kinetic energy, red line the kinetic energy in the $x_1$ direction and blue line the kinetic energy in the $x_3$ direction. Left panel is for $V_A=0.5$ and the oscillation period is $7.12$. Right panel is for $V_A=1$ and the oscillation period is $3.26$. Time is truncated at $t=20$ for better view.}\label{fig:mhd3}
\end{figure}

Similar to the inertial wave in the rotating flow, the Alfv\'en wave exists in MHD flow. At the lowest $Rm=0.1$ Alfv\'en wave cannot be found because the induction effect is too weak. At the higher $Rm$'s Alfv\'en wave is induced. Figure \ref{fig:mhd3} shows the Alfven wave at the highest $Rm=100$. The period of oscillations at $V_A=0.5$ is nearly twice of that at $V_A=1$, which is consistent with the fact that the frequency of Alfv\'en wave is proportional to the strength of imposed field. The reason that it is not exactly twice is due to the wave number $k_3$. The dimensionless frequency of Alfv\'en wave is $Vak_3$ which depends not only on $V_A$ but also on wave number $k_3$ (for comparison, the frequency of inertial wave does not depend on wave number but the orientation of wave vector), and different $V_A$ leads to different $k_3$ because of different anisotropy.

\section{Rotating MHD flow}\label{sec:rot-mhd}

In this section we study decay of rotating MHD flow. We impose a magnetic field which is static in the rotating frame of reference, namely the imposed field is co-rotating with the mean flow. This setup is often used in the study of planetary and stellar interiors, e.g. the dynamo action. In this case the angle between rotational and magnetic axes should be considered. Suppose that rotation is along the $x_3$ axis and imposed field is in the $x_1-x_3$ plane. Denote the angle between rotation and imposed field by $\alpha$ and then the dimensionless imposed field is expressed as
\begin{equation}
\bm B_0=(\sin\alpha,0,\cos\alpha).
\end{equation}
The dimensionless Navier-Stokes and induction equations are respectively
\begin{equation}
\frac{\partial\bm u}{\partial t}+\bm u\cdot\bm\nabla\bm u=-\bm\nabla p+\frac{1}{Re}\nabla^2\bm u+\frac{1}{Ro}2\bm u\times\hat{\bm x}_3+V_A^2(\bm\nabla\times\bm b)\times(\bm B_0+\bm b),
\end{equation}
\begin{equation}
\frac{\partial\bm b}{\partial t}=\bm\nabla\times\left(\bm u\times(\bm B_0+\bm b)\right)+\frac{1}{Rm}\nabla^2\bm b.
\end{equation}
One may question the validity of magnetic induction equation in the rotating frame of reference. It should be noted that the displacement current is neglected in the MHD approximation because the rotational speed of fluid is much less than the speed of light, such that the magnetic induction equation still holds in the rotating frame.

We introduce the Elsasser number measuring the ratio of the Lorentz force to the Coriolis force,
\begin{equation}
\Lambda=\frac{B_0^2}{\rho\mu\eta\Omega}.
\end{equation}
The Elsasser number plays an important role in the dynamo action. In the magnetostrophic balance, the pressure gradient, the Coriolis force and the Lorentz force are balanced. In this situation the Elsasser number is of the order of unity. It might happen in the fluid core of the Earth where the toroidal field created by the differential rotation (the so-called $\omega$ effect) is so strong that the magnetostrophic balance is reached. The Elsasser number is related to the Rossby number, the magnetic Reynolds number and the dimensionless Alvf\'en speed through the expression
\begin{equation}
\Lambda=RoRmV_A^2.
\end{equation}

\begin{figure}
\centering
\includegraphics[scale=0.7]{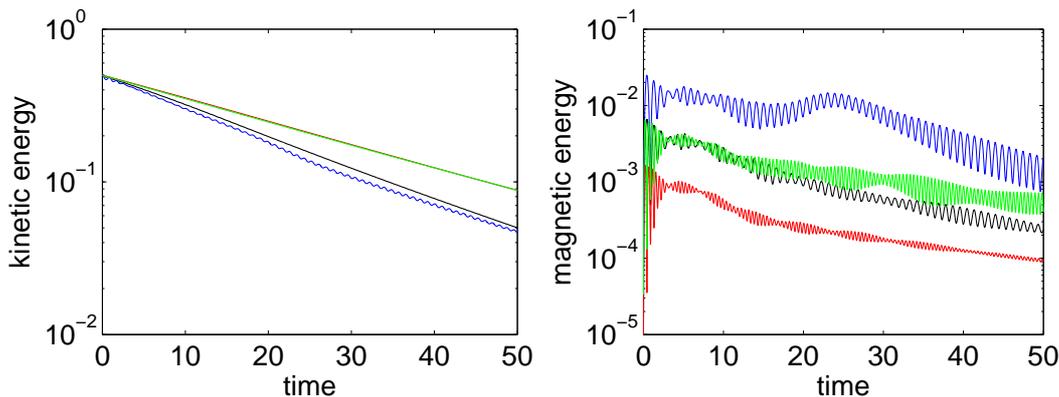}
\caption{Rotating MHD flow. The time evolution of volume-averaged kinetic energy and magnetic energy of induced field. $Rm=100$ and $\alpha=0^\circ$. Black, red, blue and green lines denote the pair of $(Ro,V_A)$ to be respectively $(0.2,0.5)$, $(0.1,0.5)$, $(0.2,1.0)$ and $(0.1,1.0)$. Time is truncated at $t=50$ for better view.}\label{fig:rot-mhd1}
\end{figure}

Through the study in \S\ref{sec:mhd} we know that low $Rm$ and high $Rm$ regimes are quite different for dynamics, namely Alfv\'en wave can be induced at high $Rm$ but is absent at low $Rm$. In geophysical and astrophysical flows, $Rm$ is high, e.g. in the Earth's fluid core $Rm$ is of the order of $100$. Therefore, we keep $Rm=100$ to study the high $Rm$ regime. $\alpha$ is kept to be $0^\circ$. We calculate the four rotating MHD flows with the combination of $Ro=(0.2, 0.1)$ and $V_A=(0.5, 1.0)$ for comparison between slow and fast rotation as well as weak and strong imposed field. Figure \ref{fig:rot-mhd1} shows the time evolution of volume-averaged kinetic energy and magnetic energy of induced field. In the left panel, the two flows at $Ro=0.1$ (red and green lines, they are almost overlapped) decay more slowly than the two flows at $Ro=0.2$ (black and blue lines), because fast rotation retards the flow decay. The fact that the flows are grouped by $Ro$ other than $V_A$ indicates that rotation wins out imposed field in the parameter regime investigated. At $Ro=0.2$ of slow rotation, the flow with stronger imposed field (blue line) decays faster than with weaker imposed field (black line) because of energy conversion studied in \S\ref{sec:mhd}. In the right panel, it is not surprising that the two flows with stronger imposed field (blue and green lines) have higher magnetic energy of induced field than the two flows with weaker imposed field (black and red lines). However, it is interesting that at each $V_A$ the flow at $Ro=0.2$ has higher magnetic energy of induced field than at $Ro=0.1$ (black versus red, blue versus green). Again, it is because of energy conversion. As indicated by the left panel, the two flows at $Ro=0.1$ have higher kinetic energy than at $Ro=0.2$. Therefore, slower rotation in rotating MHD flow tends to convert more kinetic energy to magnetic energy.

\begin{figure}
\centering
\includegraphics[scale=0.7]{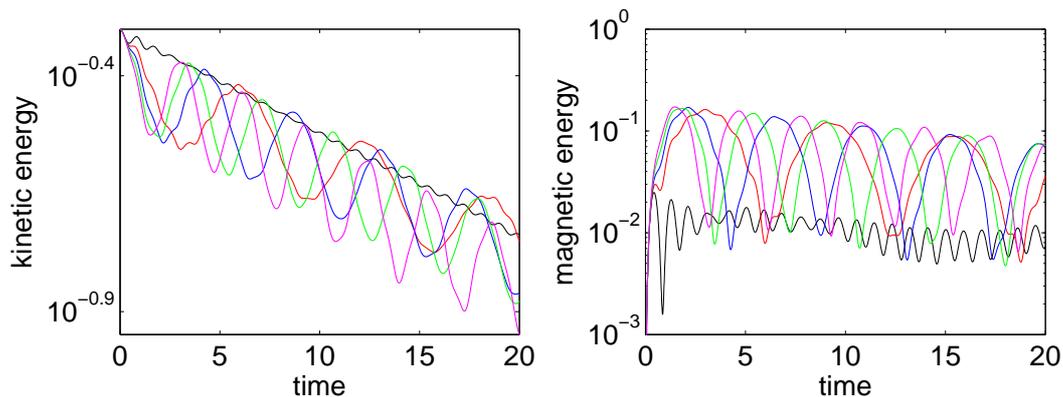}
\caption{Rotating MHD flow. The time evolution of volume-averaged kinetic energy and magnetic energy of induced field. $Ro=0.2$, $V_A=1$ and $Rm=100$. Black, red, blue, green and magenta lines denote respectively $\alpha=0^\circ$, $30^\circ$, $45^\circ$, $60^\circ$ and $90^\circ$. Time is truncated at $t=20$ for better view.}\label{fig:rot-mhd2}
\end{figure}

To end this section we study the effect of angle $\alpha$ on rotating MHD flow. We keep $Ro=0.2$, $V_A=1$ and $Rm=100$, and vary $\alpha$ from $0^\circ$ to $90^\circ$. Figure \ref{fig:rot-mhd2} shows the time evolution of volume-averaged kinetic energy and magnetic energy of induced field. Both left and right panels reveal that a noticeable difference with respect to amplitude and frequency of oscillations exists between $\alpha=0^\circ$ and $\alpha\neq0^\circ$ but this difference among the non-zero angles is not very noticeable (the phase difference always exists for different angles). More kinetic energy at $\alpha\neq0^\circ$ is converted to magnetic energy than at $\alpha=0^\circ$. Heavier oscillations at $\alpha\neq0^\circ$ than at $\alpha=0^\circ$ indicate that stronger wave is induced at $\alpha\neq0^\circ$. Therefore, a field component perpendicular to rotation causes more efficient energy conversion and induces stronger wave. This may be tentatively interpreted with the electromotive force (e.m.f.) on its first order, $\bm u\times\bm B_0$. In the presence of rotation, flow in the rotational direction wins out the other two directions, as suggested by figure \ref{fig:rot3}, such that the major contribution to e.m.f. is $u_3B_0\sin\alpha\hat{\bm x}_2$. The e.m.f. measures the interaction of flow and field. With $\alpha$ increasing, this interaction increases such that more energy conversion occurs and stronger wave is induced.

\section{Discussion}\label{sec:diss}

In this paper, we study the decay of isotropic and anisotropic flows in the weakly nonlinear regime. The parameters used in the numerical calculations are listed in Table \ref{tab:parameters}. The decay of isotropic flow has two stages, the first stage for development of small scales and the second stage for viscous dissipation. Rotation induces inertial wave and causes the formation of large-scale structure and retards the flow decay. It takes the effect in the first stage. Imposed field also causes the large-scale structure, but facilitates the flow decay in the first stage because of energy conversion from flow to magnetic field. The high and low $Rm$ regimes have different dynamics: in the former Alfv\'en wave is induced but in the latter it cannot. In the presence of both rotation and magnetic field, slower rotation tends to convert more kinetic energy to magnetic energy, and the orientation of rotation and field is important for the dynamics, namely a non-zero angle between rotation and magnetic field causes more efficient energy conversion and stronger wave than the zero angle. It is found that these isotropic and anisotropic flows in the weakly nonlinear regime exhibit the similar behaviour to the turbulent flows. Therefore, it provides some implications for the study of turbulent flow, e.g. for the decay of rotating MHD turbulent flow, the different angles between rotational and magnetic axes should be more thoroughly studied.

\begin{table}\label{tab:parameters}
\centering
\begin{tabular}{|ll|}
\hline
\multicolumn{2}{|c|}{isotropic} \\
$Re$ & $k_0$ \\
100, 200 & 1, 2 \\
\hline
\end{tabular}
\begin{tabular}{|lll|}
\hline
\multicolumn{3}{|c|}{rotating} \\
$Re$ & $Ro$ & $k_0$ \\
100 & 1.0, 0.5, 0.2, 0.1 & 1 \\
\hline
\end{tabular}
\begin{tabular}{|llll|}
\hline
\multicolumn{4}{|c|}{MHD} \\
$Re$ & $V_A$ & $Rm$ & $k_0$ \\
100 & 0.5, 1, 2, 5 & 0.1, 1, 10, 100 & 1 \\
\hline
\end{tabular} \\
\begin{tabular}{|lllll|}
\hline
\multicolumn{5}{|c|}{rotating MHD} \\
$Re$ & $(Ro, V_A)$ & $Rm$ & $\alpha$ & $k_0$ \\
100 & (0.2, 0.5), (0.1, 0.5), (0.2, 1.0), (0.1, 1.0) & 100 & 0$^\circ$, 30$^\circ$, 45$^\circ$, 60$^\circ$, 90$^\circ$ & 1 \\
\hline
\end{tabular}
\caption{The parameters used in the numerical calculations for different types of flows.}
\end{table}

\begin{acknowledgements}
This work was started in Princeton and completed in Shanghai. This work was partly supported by the National Science Foundation’s Center for Magnetic Self-Organization under grant PHY-0821899 and partly by the start-up grant of Shanghai Jiao Tong University.
\end{acknowledgements}

\bibliographystyle{spbasic}
\bibliography{paper}

\end{document}